\documentclass[10pt,a4paper,english]{revtex4}
\usepackage[T1]{fontenc}
\usepackage[latin1]{inputenc}

\makeatletter

\providecommand{\LyX}{L\kern-.1667em\lower.25em\hbox{Y}\kern-.125emX\@}
\usepackage{graphicx}

\usepackage{babel}
\makeatother
\begin{document}

\title{N--body simulations for coupled dark energy: halo mass function and
density profiles}

\author{A.V. Macci\`{o}$^{1}$, C. Quercellini$^{2,3}$, R. Mainini$^{1}$,
L. Amendola$^{2}$ \& S.A. Bonometto$^{1}$}

\affiliation{$^{1}$Universit\`{a} di Milano Bicocca, Piazza della Scienza 3,
20126 Milano, Italy ~\&~ I.N.F.N., Sezione di Milano}

\affiliation{$^{2}$INAF/ Osservatorio Astronomico di Roma, Via Frascati 33, 00040
Monte Porzio Catone, Italy }

\affiliation{$^{3}$Universit\`{a} di Roma - Tor Vergata, Via della Ricerca Scientifica
1, 00133 Roma - Italy}

\begin{abstract}
We present the results of a series of $N$-body simulations in cosmologies
where dark matter (DM) is coupled to dark energy (DE), so easing the
cosmic coincidence problem. The dark--dark coupling introduces two
novel effects in N--body dynamics: (i) DM particle masses vary with
time; (ii) gravity between DM particles is ruled by a constant $G^{*}$,
greater than Newton's constant $G$, holding in other 2--body interactions.
As a consequence, baryons and DM particle distributions develop a
large scale bias. Here we investigate DE models with Ratra-Peebles
(RP) potentials; the dark--dark coupling is set in a parametric range
compatible with observations, for as concern background and linear
perturbation properties. We study the halo mass function, the halo
density profile and the behavior of the non-linear bias. We find that
non-linear dynamics puts additional constraints to the coupling parameter.
They mostly arise from density profiles, that we find to yield higher
concentrations, in coupled RP models, with respect to (uncoupled)
dynamical DE cosmologies. Such enhancement, although being a strong
effect in some coupling parameter range, is just a minor change for
smaller but significant values of the coupling parameter. With these
further restrictions, coupled DE models with RP potential are consistent
with non--linear observables. 
\end{abstract}
\maketitle

\section{Introduction}

\label{sec:intro}

The nature of Dark Energy (DE) is one of the main puzzles of cosmology.
DE was first required by SNIa data \cite{snia}, but a \textit{flat}
Universe with $\Omega _{m}\simeq 0.3$, $h\simeq 0.7$ and $\Omega _{b}h^{2}\simeq 0.02$
is also favored by CMB and LSS observations \cite{wmap} ($\Omega _{m,b}$:
matter, baryon density parameters; $h$: Hubble parameter in units
of 100 km/s/Mpc; CMB: cosmic microwave background radiation; LSS:
large scale structure). 

DE could be a false vacuum; from the expression of its stress--energy
tensor, $T_{\mu \nu }^{(DE)}=\Lambda g_{\mu \nu }$ ($\Lambda $ is
a positive constant and $g_{\mu \nu }$ is the metric tensor), one
immediately appreciates that its pressure and energy density ($p_{DE}$
and $\rho _{DE}$) have ratio $w=-1$. False vacuum, however, requires
a severe fine tuning at the end of the EW transition. Otherwise, DE
could be a scalar field $\phi $ self--interacting through a potential
$V(\phi )$. Then \begin{equation}
\rho _{DE}={{\dot{\phi }}^{2}/2a^{2}}+V(\phi ),\qquad p_{DE}={{\dot{\phi }}^{2}/2a^{2}}-V(\phi )\end{equation}
 (here dots indicate differentiation in respect to the conformal time
$\tau $). However, if the kinetic component $\rho _{k}(\phi )\equiv {\dot{\phi }}^{2}/2a^{2}$
approaches the potential component $V(\phi )$, $p_{DE}$ vanishes
and the scalar field $\phi $ behaves as CDM (Cold Dark Matter). This
happens, e.g., in the well known case of axion DM. But, even for lower
$\rho _{k}$, if $1/2<\rho _{k}/V<1$, it is $-1/3<w<0$ and the model,
at most, approaches an open CDM behavior. The relevant domain is attained
when $\rho _{k}/V\ll 1/2$, although keeping a state parameter $w>-1$.
Then, $\phi $ approaches a $\Lambda $CDM behavior and is currently
dubbed \textit{dynamical} DE \cite{amen02} \cite{am-mnras} \cite{stei99}. 

The conceptual contiguity between DM and dynamical DE suggests that
they may not be disjoint entities. If so, one could hopefully ease
the cosmic coincidence problem, i.e. that DM and DE densities, after
being different by orders of magnitude for most of the cosmic history,
approach equal values only in today's Universe. The simplest way to
deal with this idea amounts to admit an interaction between DM and
DE \cite{amen00}. In a number of papers \cite{grad92,amen02,am-mnras,amed02},
it has been shown that, at the linear level, this causes no apparent
conflict with LSS or CBR data. We shall refer to models where DM and
dynamical DE interact as \textit{coupled} DE models. Other models that
propose a direct
link between DM and DE invoke a unified model (e.g. \cite{bertolami}) or condensation mechanisms \cite{bassett}.

An open question, then, concerns the emergence of non--linear structures
in these models and how easy it is to fit observed LSS data with predictable
features. In this work we deal with this problem by using N--body
simulations of coupled DE models for DE self--interacting through
a Ratra--Peebles \cite{rape88}{[} RP hereafter{]} potential: \begin{equation}
V(\phi )=\Lambda ^{4+\alpha }/\phi ^{\alpha }.\end{equation}
 Once the exponent $\alpha $ and the DE density parameter $\Omega _{DE}$
are fixed, the energy scale $\Lambda $ is set. This self--interaction
allows $w\ll -1/3$, if $\Lambda $ is sufficiently low. 

Let us outline soon that our non--linear treatment sets precise constraints
to coupled DE parameters. A wide parameter space however remains where,
apparently, these models fit LSS data as well as those with (uncoupled)
dynamical DE, although coupled DE, with RP potential, allows no improvement
of such fit. The above motivations for coupled DE \cite{damo90} however
remain and, altogether, the non--linear test is successfully passed.
N--body simulations of models with (uncoupled) dynamical DE were recently
performed \cite{mmbk} \cite{kmmb} \cite{lin}. Here we follow the
same pattern of \cite{kmmb} and use the program ART \cite{art} providing,
first of all, the fair dependence of the matter density parameter
$\Omega _{m}$ on the scale factor $a$. To our knowledge, this is
the first time an $N$-body simulation with species-dependent scalar
gravity is carried out. Our conclusions are based on simulations of
a variety of models with different RP slopes $\alpha $ and coupling
parameters $\beta $. Let us list them soon: first of all, we test
two $\alpha $ slopes: 0.143 and 2. The latter value approaches the
greatest value for which agreement with CMB observations is granted
\cite{amen02}. This is the range of RP models which are most distant
from $\Lambda $CDM. We explored also a wide set of $\beta $ couplings,
ranging from $0.05$ to $0.25$. All simulations were performed starting
from the same random numbers and, for the sake of comparison, we also
run a $\Lambda $CDM simulation starting from such random numbers.
The other parameters were set to values chosen in agreement with recent
CMB experiments \cite{wmap}: $\Omega _{c}h^{2}=0.15$, $\Omega _{b}h^{2}=0.01$,
$h=0.7$ ($\Omega _{c}$ is the (cold) DM density parameter). All
models are normalized so that $\sigma _{8}=0.75$ today, to match
both CMB data and the observed cluster abundance \cite{sch}. Further
details on the simulation performed are listed in Table 1. 

\begin{center}\begin{tabular}{|c|c|c|c|c|c|c|}
\hline 
Model&
 $\alpha $&
 $\beta $&
 Box size ($h^{-1}$Mpc)&
 \# of particle&
 Mass res. ($h^{-1}$M$_{\odot })$&
 Force res. ($h^{-1}$kpc)\\
\hline
RP$_{1}$&
 0.143&
 0.05&
 80&
 128$^{3}$&
 2.0$\times $10$^{10}$&
 5\\
\hline
RP$_{2}$&
 0.143&
 0.10&
 80&
 128$^{3}$&
 2.0$\times $10$^{10}$&
 5\\
\hline
RP$_{3}$&
 0.143&
 0.15&
 80&
 128$^{3}$&
 2.0$\times $10$^{10}$&
 5\\
\hline
RP$_{4}$&
 0.143&
 0.2&
 80&
 128$^{3}$&
 2.0$\times $10$^{10}$&
 1.2\\
\hline
RP$_{5}$&
 0.143&
 0.25&
 80&
 128$^{3}$&
 2.0$\times $10$^{10}$&
 1.2\\
\hline
RP$_{6}$&
 2.0&
 0.15&
 80&
 128$^{3}$&
 2.0$\times $10$^{10}$&
 5\\
\hline
$\Lambda $CDM&
 0&
 0&
 80&
 128$^{3}$&
 2.0$\times $10$^{10}$&
 5\\
\hline
\multicolumn{7}{|c|}{Table I. }\\
\hline
\end{tabular}\end{center}

For all these models we also run an high-resolution simulation of
a single halo, with a mass resolution of $2.5\times 10^{9}h^{-1}$M$_{\odot }$
and a force resoltuion of $1.2h^{-1}$kpc. 

In the next section we discuss the linear and post--linear aspects
of coupled DE, explaining, in particular, how $\Omega _{m}(a)$ is
obtained and used. In section 3 we focus on the newtonian regime for
coupled DE models and describe the different gravitation of baryons
and DM. In Section 4, we implement these prescriptions in the numerical
code, so explaining which further modifications ART needs, to deal
with coupled DE. Section 5 is then devoted to illustrating the results
while, in Section 6, we draw our conclusions.

\section{Background expansion in models with coupled dark energy}

Quite in general, energy density and pressure of each component, in
models with dynamical DE, are obtainable from the stress--energy tensors
$T_{\mu \nu }^{(c,b,r,\phi )}$, (for CDM, baryons, radiation, and
DE, respectively; radiation includes neutrinos). General covariance
requires that the sum $T_{\mu \nu }$ of these four tensors fulfills
the \textit{continuity equation}\begin{equation}
T_{\nu ;\mu }^{\mu }=0\label{continu}\end{equation}
 and, although this is true if all tensors fulfill it separately,
such requirement is not necessary; e.g., when we deal with fluctuations
before hydrogen recombination, only the sum of baryon and e.m. radiation
tensors fulfills it. In a similar way, if DE and DM interact, we can
have that \begin{eqnarray}
T_{\nu ;\mu }^{(\phi )\, \mu } & = & \sqrt{\frac{16\pi G}{3}}\beta T^{(c)}\phi _{;\nu }\label{coupling}\\
T_{\nu ;\mu }^{(c)\, \mu } & = & -\sqrt{\frac{16\pi G}{3}}\beta T^{(c)}\phi _{;\nu }\nonumber 
\end{eqnarray}
 (here $T^{(c)}$ is the trace of the CDM stress--energy tensor),
and the sum of DM and DE fulfills eq.~(\ref{continu}). No analogous
interaction should involve baryons, for which we assume that $T_{\nu ;\mu }^{(b)\, \mu }=0$;
in fact, experimental and observational constraints restrict an hypotetical
DE--baryon coupling to $\beta _{b}<0.01$ \cite{damo93}; also radiation
cannot be involved, as the trace of its stress-energy tensor vanishes
because of its equation of state. The particular form of the coupling
(\ref{coupling}) reduces to Brans-Dicke scalar gravity upon a conformal
transformation (see, e.g. \cite{damo93,wett95,damo90,amen99}). 

We assume a flat conformal \textit{background} metric $ds^{2}=a^{2}(-d\tau ^{2}+\delta _{ij}dx^{i}dx^{j})$
($i,j=1,..,3$), so that the background continuity equations read
\begin{eqnarray}
\ddot{\phi }+2{\mathcal{H}}\dot{\phi }+a^{2}V_{,\phi } & = & \sqrt{16\pi G/3}\beta a^{2}\rho _{c},\label{continu1}\\
\dot{\rho _{c}}+3{\mathcal{H}}\rho _{c} & = & -\sqrt{16\pi G/3}\beta \rho _{c}\dot{\phi },\\
\dot{\rho _{b}}+3{\mathcal{H}}\rho _{b} & = & 0,\\
\dot{\rho _{r}}+4{\mathcal{H}}\rho _{r} & = & 0.
\end{eqnarray}
 being ${\mathcal{H}}=\dot{a}/a$ the conformal Hubble parameter.
The dimensionless constant $\beta ^{2}$ can be seen as the ratio
of the DE--DM interaction with respect to gravity \cite{wett95,amen00}.
In these models, after \textit{equivalence}, the world passes through
three different expansion regimes, denominated: (i) $\phi $--MDE,
(ii) tracking phase (\cite{stei99}), and a final (iii) global attractor. 

Immediately after equivalence, the world enters a $\phi $--MDE stage
($\phi $--matter dominated expansion, see \cite{amen00}), not far
from a matter dominated expansion (MDE), although one should not neglect
a small correction, proportional to $\beta $, due to the kinetical
term $\rho _{k}(\phi )$. $V(\phi )$, instead, is negligible and,
accordingly, the correction is indipendent of the potential shape.
By solving the Friedmann equation we find that \begin{equation}
a\propto t^{4/(6+4\beta ^{2})},\label{scale}\end{equation}
 i.e. that the scale factor grows more slowly than in a pure MDE (in
Section V.B, we shall see that, on the contrary, the perturbation
growth is enhanced, during this stage). During the $\phi $--MDE stage,
$V(\phi )$ gradually increases and, eventually, approaches and exceeds
$\rho _{k}(\phi )$; then, the world enters the \textit{tracking phase},
whose details depend on the potential shape; for most potentials,
such phase ends up into a \textit{global attractor}, when DE density
overwhelms DM and any other densities. 

Along the expansion history, the scaling of $\rho _{c}$ (DM density)
is modified with respect to the uncoupled case and reads \begin{equation}
\rho _{c}={\frac{\rho _{oc}}{a^{3}}}e^{-\sqrt{\frac{16\pi G}{3}}\beta (\phi -\phi _{0})};\label{densi}\end{equation}
 here the subscript $o$ indicates value at the present time $\tau _{o}$,
while we take $a_{o}=1$. Meanwhile the baryon density grows as $a^{-3}$,
as usual. 

\begin{figure*}
\includegraphics[  scale=0.3]{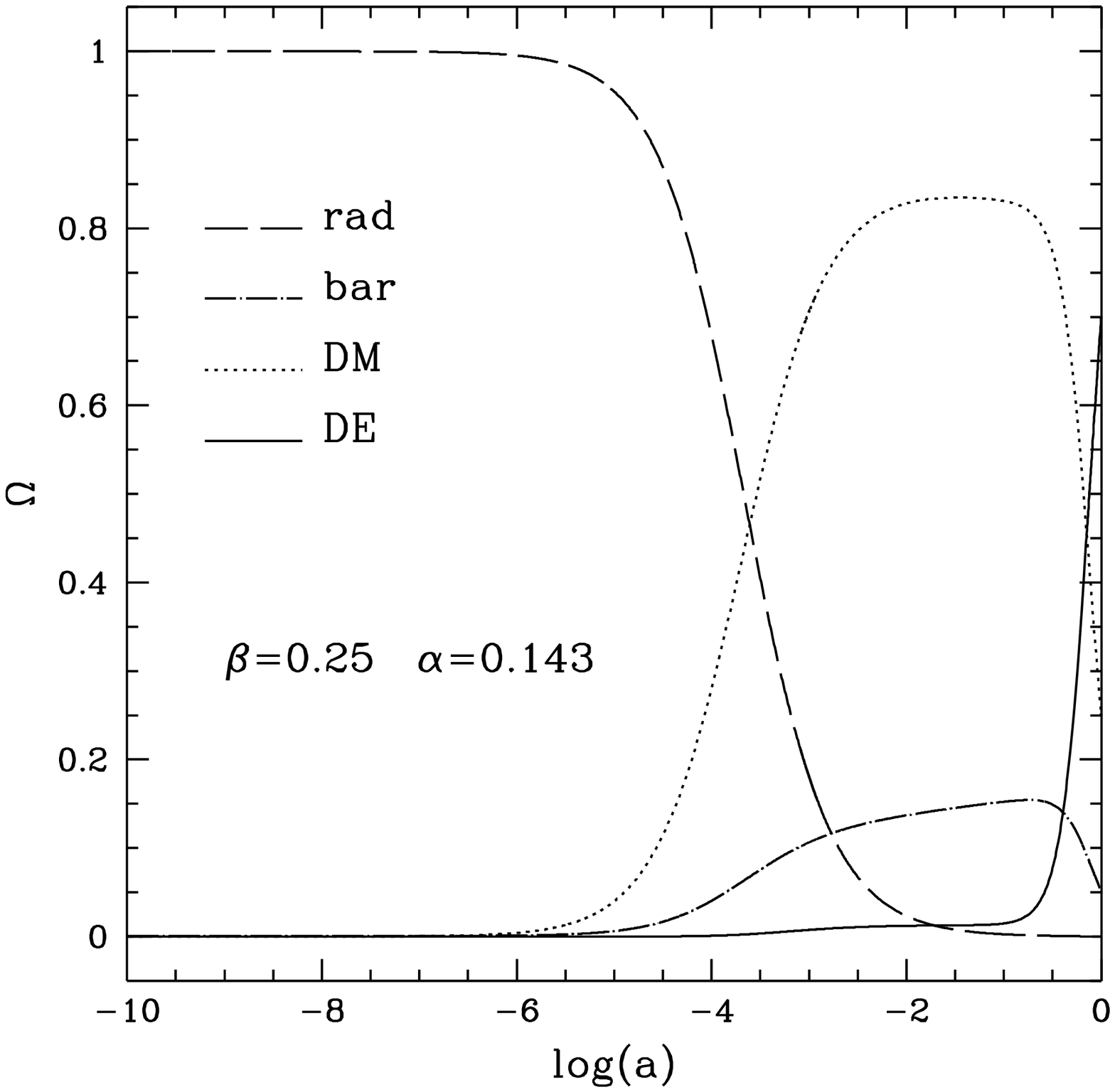}
 \includegraphics[  scale=0.3] {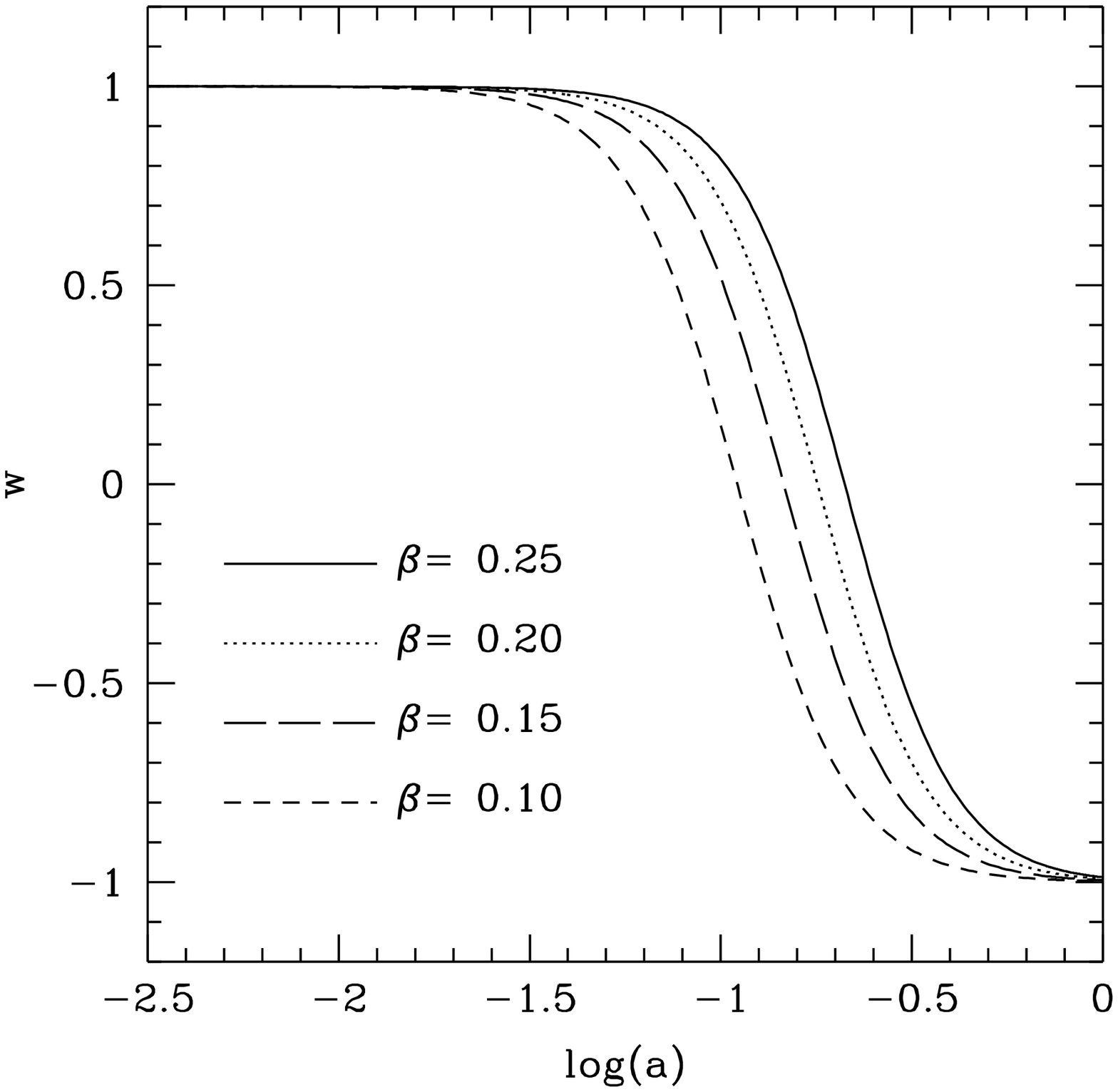}

\caption{Left panel: density parameters for radiation, baryons, DM and DE.
Just after radiation equivalence, a DE plateau occurs, due to the
dark--dark coupling. In the text this plateau is denominated $\phi $--MDE
stage. ~Right panel: evolution of the state parameter $w$ for different
value of $\beta $}

\label{fig:den}
\end{figure*}

In the next section we shall see that these behaviors strongly affect
the fluctuation growth, even in the newtonian regime. Fig. \ref{fig:den}
(left panel) shows the different trends of the density parameters
for two different values of the coupling parameter $\beta $. The
three stages of the background evolution are clearly visible. For
sake of completness we also report in Fig.~\ref{fig:den} (right
panel) the $a$ dependence of the state parameter $w$ for different
value of $\beta $. Notice that, for RP models with a low $\alpha $,
the value of the state parameter has been quite close to $-1$, since
a redshift $\sim 3$--4. 

All along these stages, however, according to the Friedman equations,
the following relation holds \cite{mmbk}: \begin{equation}
{\frac{dt}{da}}\equiv \chi (a)=H_{o}^{-1}\sqrt{\frac{a\Omega _{b}(a)}{\Omega _{b}(a_{o})}}.\label{chi}\end{equation}
 Therefore, once the $a$ dependence of the baryon density parameter
is given, all derivatives in respect to time can be easily converted
into derivatives in respect to the scale factor. This relation will
be used in the implementation of the ART program. 

\section{Dynamics in the newtonian regime}

\label{dyn} Let us now consider density fluctuations and discuss
their evolution. First of all, the conformal metric must be modified
to take into account the local gravitational fields and, in the absence
of anisotropic stresses, reads \[
ds^{2}=a^{2}[-(1+2\Phi )d\tau ^{2}+(1-2\Phi )dx_{i}dx^{j}],\]
 $\Phi $ being the gravitational potential. Let us use $\alpha =\log a$
as independent variable, instead of $\tau $. Differentiation in respect
to $\alpha $ will be indicated by '. As usual, fluctuations are expanded
in Fourier components; let us then consider a component of wavenumber
$k$ and define $\lambda ={\mathcal{H}}/k$. 

Baryons and CDM will be considered as fluids; fluctuations will then
be characterized by density fluctuations \begin{equation}
\delta ^{c,b}=\delta \rho _{c,b}/\rho _{c,b}\label{delta}\end{equation}
 ($^{c,b}$ stand for CDM and bayons) and velocity fields $v_{i}^{c,b}\equiv dx_{i}/d\tau $,
from which we build the scalar variables \begin{equation}
\theta ^{c,b}=i{\frac{k_{i}{v^{c,b}}_{i}}{{\mathcal{H}}}}.\label{theta}\end{equation}
 Scalar field fluctuations ($\delta \phi $) will then be described
by the variable \begin{equation}
\varphi =\sqrt{\frac{3}{4\pi G}}\, \delta \phi .\label{varphi}\end{equation}
 Dealing with fluctuation evolution well after recombination, we shall
neglect their radiation component. 

Taking into account density inhomogeneities, the equations (\ref{continu})
yield the dependence of $\delta ^{c,b}$ and $\theta ^{c,b}$ on the
scale factor $a$. The Friedman equation shall then be used to obtain
the dependence on time. We shall omit here the general form of these
equations (see \cite{amed02}), and will write them in the Newtonian
limit, i.e. for small scales (in comparison with the horizon scale
$\sim H^{-1}$) and small velocities (in respect to $c$). 

The former condition tells us to consider the lowest order terms in
$\lambda $; in this limit, the gravitational potential fulfills equations
that can be written in a simpler form by defining the function $f(\phi )$
according to \begin{equation}
V(\phi )=A\exp \big [\sqrt{16\pi G/3}f(\phi )\phi \big ],\end{equation}
 as well as the functions \begin{equation}
f_{1}=\phi {\frac{df}{d\phi }}+f,\quad f_{2}=\phi {\frac{df}{d\phi }}+2f+f_{1};\end{equation}
 notice that this is no restriction on the potential shape. Then \begin{eqnarray}
 & \Phi  & =-\frac{3}{2}\lambda ^{2}(\Omega _{b}\delta ^{b}+\Omega _{c}\delta ^{c}+6X\varphi +2X\varphi '-2Y^{2}f_{1}\varphi )\label{poten1}\\
 & \Phi ' & =3X\varphi -\Phi ,\label{poten2}
\end{eqnarray}
 while the scalar field fulfills the equation \begin{equation}
\varphi ''+\big (2+\frac{{\mathcal{H}}^{'}}{{\mathcal{H}}}\big )\varphi '+\lambda ^{-2}\varphi -12X\varphi +4\Phi X+2Y^{2}(f_{2}\varphi -f_{1}\Phi )=\beta \Omega _{c}(\delta ^{c}+2\Phi ).\label{phi2}\end{equation}
 These expressions have been simplified by using the variables $X^{2}=8\pi G\rho _{k}(\phi )\, a^{2}/3{\mathcal{H}}^{2}$
(kinetic density parameter) and $Y^{2}=8\pi GV(\phi )\, a^{2}/3{\mathcal{H}}^{2}$
(potential densiity parameter) . It follows that if the DE kinetic
(or potential) energy density gives a substantial contribution to
the expansion source, $X$ (or $Y$) is $\mathcal{O}$(1). 

In the Newtonian limit, however, we also neglect the derivatives of
$\varphi $, averaging out the rapid oscillations of $\varphi $ and
the potential term $f_{2}Y^{2}\varphi $; this actually requires that
$\lambda <<(f_{2}Y)^{-1}$ (remind that $Y$ is $\mathcal{O}$(1)).
Furthermore, in eq.~(\ref{phi2}), the metric potential $\Phi $,
which is proportional to $\lambda ^{2}$, can also be neglected. Accordingly,
eq.~(\ref{poten1}) becomes \begin{equation}
\Phi =-\frac{3}{2}\lambda ^{2}(\Omega _{b}\delta ^{b}+\Omega _{c}\delta ^{c}),\
 \label{approx1}\end{equation}
 which is the usual Poisson equation, while eq.~(\ref{phi2}) simplifies
into \begin{equation}
\lambda ^{-2}\varphi \simeq \beta \Omega _{c}\delta ^{c}.\label{approx2}\end{equation}
 In the same way (see \cite{amed02}), we obtain the continuity equations
\begin{eqnarray}
{\delta ^{c}}\, '' & = & -{\delta ^{c}}\, '\big (1+\frac{{\mathcal{H}}'}{\mathcal{H}}-2\beta X\big )+\frac{3}{2}(1+\frac{4}{3}\beta ^{2})\Omega _{c}\delta ^{c}+\frac{3}{2}\Omega _{b}\delta ^{b},\label{delta2}\\
{\delta ^{b}}\, '' & = & -{\delta ^{b}}\, '\big (1+\frac{{\mathcal{H}}'}{\mathcal{H}}\big )+\frac{3}{2}(\Omega _{c}\delta ^{c}+\Omega _{b}\delta ^{b}),\label{delta2b}
\end{eqnarray}
 and the Euler equations \begin{eqnarray}
{\theta ^{c}}\, ' & = & -\theta ^{c}\big (1+\frac{{\mathcal{H}}'}{\mathcal{H}}-2\beta X\big )-\frac{3}{2}\big (1+\frac{4}{3}\beta ^{2}\big )\Omega _{c}\delta ^{c}-\frac{3}{2}\Omega _{b}\delta ^{b},\label{velocity}\\
{\theta ^{b}}\, ' & = & -\theta ^{b}\big (1+\frac{{\mathcal{H}}'}{\mathcal{H}}\big )-\frac{3}{2}(\Omega _{c}\delta ^{c}+\Omega _{b}\delta ^{b}).\label{velocityb}
\end{eqnarray}
 From the latter ones, we can derive the acceleration of a single
non-relativistic CDM or baryon particle (mass $m_{c,b}$), assuming
that it lays in the empty space, at distance $r$ from the origin,
where either a CDM particle of mass $M_{c}$ or a baryon particle
of mass $M_{b}$ are set. 

In fact, owing to eq.~(\ref{densi}), normalizing the scalar field
so that its present value $\phi _{o}=0$, and assuming that the density
of the particle is much larger than the background density, it shall
be \begin{eqnarray*}
\Omega _{c}\delta ^{c} & = & \frac{\rho _{M_{c}}-\rho _{c}}{\rho _{crit}}=\frac{8\pi Ge^{-\sqrt{\frac{16\pi G}{3}}\, \beta \phi }M_{c}\delta (0)}{3{\mathcal{H}}^{2}a}\\
\Omega _{b}\delta ^{b} & = & \frac{\rho _{M_{b}}-\rho _{b}}{\rho _{crit}}=\frac{8\pi G\delta (0)}{3{\mathcal{H}}^{2}a},
\end{eqnarray*}
 $\delta $ being the Dirac distribution. Then, reminding that the
divergence $\nabla _{i}v_{i}^{c,b}=\theta ^{c,b}{\mathcal{H}}$, and
using the ordinary (not conformal) time variable, instead of $\alpha $,
eq.~(\ref{velocity}) yields \begin{equation}
{\nabla }_{i}\dot{v_{i}^{c}}=-{H}(1-2\beta X)\nabla _{i}v_{i}^{c}-\frac{4\pi G^{*}M_{c}e^{-\sqrt{\frac{16\pi G}{3}}\, \beta \phi }\delta (0)}{a^{2}}-\frac{4\pi GM_{b}\delta (0)}{a^{2}},\label{velocity2}\end{equation}
 where dots yield differentiation in respect to ordinary time, \begin{equation}
G^{*}=G(1+\frac{4}{3}\beta ^{2})\label{gi}\end{equation}
 and $H=d\log a/dt$. 

We can integrate this equation, taking into account that the acceleration
is radial, as the attracting particle lays in the origin. It will
then be \[
\int d^{3}x\, \nabla \cdot {\dot{\textbf {v}}}=4\pi \int dx\frac{d(x^{2}\dot{v})}{dx}=4\pi x^{2}\dot{v},\]
 $\dot{v}$ being the modulus of the (radial) acceleration (in the
second term $x=|{\textbf {x}}|$). Accordingly, for a CDM particle,
the desired expression of the radial acceleration reads \begin{equation}
\dot{v}^{c}=-(1-2\beta X)H{\textbf {v}}^{c}\cdot {\textbf {n}}-\frac{G^{*}M_{c}e^{-\sqrt{\frac{16\pi G}{3}}\, \beta \phi }}{r^{2}}-\frac{GM_{b}}{r^{2}},\label{finalc}\end{equation}
 ($\textbf {n}$ is a unit vector in the radial direction; $r=ax$)
which has various peculiarities and ought to be suitably commented.
To this aim it is important to compare it with the radial acceleration
\begin{equation}
\dot{v}^{b}=-H{\textbf {v}}^{b}\cdot {\textbf {n}}-{\frac{GM_{c}e^{-\sqrt{\frac{16\pi G}{3}}\, \beta \phi }}{r^{2}}}-{\frac{GM_{b}}{r^{2}}}\label{finalb}\end{equation}
 of a baryon particle. In the expression (\ref{finalc}), notice first
the velocity term. This is a peculiar acceleration that exist even
in the absence of particles displaying their attraction; its presence
means that CDM matter is not expanding in a comoving way, due to the
extra gravity it feels. Accordingly, its particles do not follow geodesics,
because their mass changes in time, and their ordinary (not comoving)
linear momentum obeys the equation \[
\dot{p_{c}}=-\frac{G^{*}M_{c}e^{-\sqrt{\frac{16\pi G}{3}}\, \beta \phi }}{r^{2}}-\frac{GM_{b}}{r^{2}}.\]

Baryon particles, instead, safely follow geodesics, although feeling
that CDM particle masses are varying. 

Let us conclude this section by summarising its specific findings:
(i) The mass assigned to CDM particles does vary in time, being $m_{c}=m_{o}e^{-\sqrt{\frac{16\pi G}{3}}\, \beta \phi }$,
while baryon particles do keep a constant mass. (ii) When interacting
between them, CDM particles feel an effective gravitational constant
$G^{*}=G(1+4\beta ^{2}/3)$; any other particle--particle interaction
occurs with the ordinary newtonian interaction constant $G$.

\section{Methods}

Particle mass variations and different interaction constants ought
to be taken into account in performing N--body simulations. They will
be based on the Adaptive Refiniment Tree code (ART) \cite{art} that
has been suitably modified to deal with coupled DE models. The ART
code starts with a uniform grid, which covers the whole computational
box. This grid defines the lowest (zeroth) level of resolution of
the simulation. The standard Particle-Mesh algorithms are used to
compute density and gravitational potential on the zero-level mesh.
The ART code reaches high force resolution by refining all high density
regions using an automated refinement algorithm. The refiniments are
recursive: the refined regions can also be refined, each subsequent
refinement meshes of different resolution, size, and geometry covering
regions of interests. Because each individual cubic cell can be refined,
the shape of the refinement mesh can be arbitrary and match effectively
the geometry of the region of interest. 

The criterion for refinement is the local density of particles: if
the number of particles ina a mesh cell (as estimated by the Cloud-In-Cell
method) exceeds the level of $n_{thresh}$, the cell is split (\char`\"{}refined\char`\"{})
in 8 cell of the next refinement level. The refinement threshold depends
on the refinement level. For the zero's level it is $n_{thresh}=2$.
For the higher level it is set to $n_{thresh}=4$. The code uses the
expansion parameter $a$ as the variable tiome. During the integration,
spatial refinement is accompanied by temporal refinement. Namely,
each level of refinement, $l$, is integrated with its own time step
$\Delta a_{l}=\Delta a_{o}/2^{l}$,where $\Delta a_{o}=3\cdot 10^{-3}$
is the global time step of the zeroth refinement level. This variable
time stepping is very important for the accuracy of the results. As
the force resolution increases, more steps are needed to integrate
the trajectories accurately. Extensive tests of the code and comparison
with other numerical N-body codes can be found in \cite{kne}. 

Let us now describe the three main modifications we made to handle
coupled DE. A first step amounts to distinguish between baryons and
DM particles, which feel different gravitational forces. Therefore,
the potential on the grid is to be calculated twice, so to fix the
different forces that baryon and DM particles feel. All particles
act on baryons through the usual gravitational constant $G$, which
sets also the action of baryons on DM particles. DM particles instead,
act on DM particles through a different interaction constant $G^{*}=G(1+4\beta ^{2}/3)$.
The gravitational force is then computed through the usual FFT approach. 

A second step amounts to take into account that the effective mass
of DM particles is time varying. Aside of the acceleration due to
gravitation, each DM particle will therefore undergo an extra acceleration
$2\beta X$. Besides of these two changes, peculiar of coupled DE
models, we ought to take into account the right relation between $a$
and $t$, as shown in eq.~(\ref{chi}), where $\chi (a)=dt/da$ is
given. By solving the background equations, in a suitable file we
provide $\chi (a)$ in $\simeq 200$ scale factor values $a_{i}$,
that we than interpolate. 

The models listed in Table 1 were first simulated in a $80h^{-1}$Mpc
box. We then selected the same halo in all simulations and magnified
it. The low--resolution simulation, performed with a force resolution
of $15\, h^{-1}$ Mpc and a mass resolution $\simeq 2\cdot 10^{10}\, h^{-1}M_{\odot }$,
allowed us to evaluate the halo mass function. The high--resulution
simulation, performed with a force resolution of $\simeq 1.2\, h^{-1}$kpc
and a mass resolution of $2.54\cdot 10^{9}h^{-1}M_{\odot }$, magnified
a sphere with a radius of $5\, h^{-1}$Mpc, centered on the halo,
allowing us to compare halo profiles down to a radius $\simeq 5\, h^{-1}$kpc. 

Besides of the above points, we could also test the non--linear evolution
of the \textit{bias} between the amplitudes of inhomogeneities in
baryons and DM. Such bias is one of the most peculiar features of
coupled DE models and we shall describe how non--linearity modifies
it.

\section{Results}

\label{results}

\subsection{Mass function}

We identify halos in simulations by using a SO algorithm, that we
shall now describe in more detail. As first step, candidate halos
are located by a FoF procedure, with linking length $\lambda =U\times d$
($d$ is the average particle separation) and keeping groups with
more than $N_{f}$ particles ($U$ and $N_{f}$ fixed herebelow).
We then perform two operations: (i) we find the point, $C_{W}$, where
the gravitational potential, due to the group of particle, has a minimum;
(ii) we determine the radius $\overline{r}$ of a sphere, centered
in $C_{W}$, where the density contrast is $\Delta _{v}$ (we use
the virial density contrast found in the absence of dark--dark coupling
\cite{mmb,mmbk}). Taking all particles within $\overline{r}$ we
perform again the operations (i) and (ii). The procedure is iterated
until we converge onto a stable particle set. The set is discarded
if, at some stage, we have less than $N_{f}$ particles. If a particle
is a potential member of two groups it is assigned to the more massive
one. In this work we use $U=0.2$ and take $N_{f}$ so to have a mass
threshold $5.0\cdot 10^{12}h^{-1}$M$_{\odot }$. 

\begin{figure}
\includegraphics[  scale=0.4]{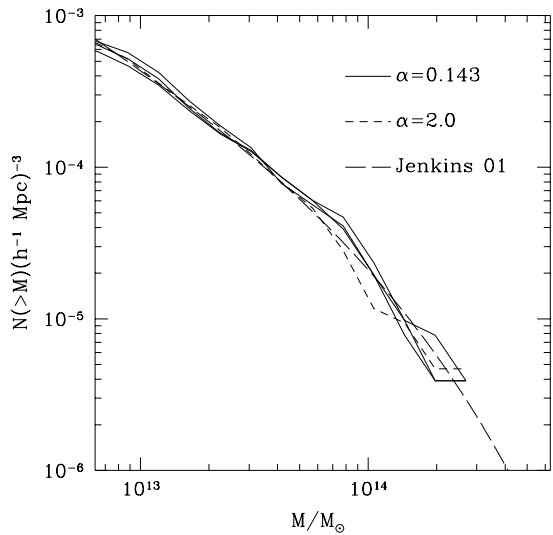}
\caption{Mass function at $z=0$ for $\alpha =2$ and $\alpha =0.143$. For
$\alpha =0.143$ we report three curves, for different values of $\beta $.
They are all practically indistinguishable and are well fitted the
approximation of Jenkins et al (2001).}

\label{fig:mf}
\end{figure}

Fig.~\ref{fig:mf} shows the mass function for isolated halos for
models with different values of $\alpha $ and $\beta $. Let us remind
that the simulations have the same initial phases and the same value
$\sigma _{8}=0.75$. Thus, the differences between models are only
due to different couplings or $w(t)$. Remarkably, at $z=0$ the mass-functions
are practically indistinguishable: a mass-function has no \char`\"{}memory\char`\"{}
of the past evolution. The mass-function obtained in this way is well
fitted by the approximation provided by \cite{jen} for $\Lambda $CDM
models (long dashed line in Fig.~\ref{fig:mf}). 

\subsection{Linear and non--linear bias}

\label{sec:linnonlin}

\begin{figure}
 \includegraphics[  scale=0.3]{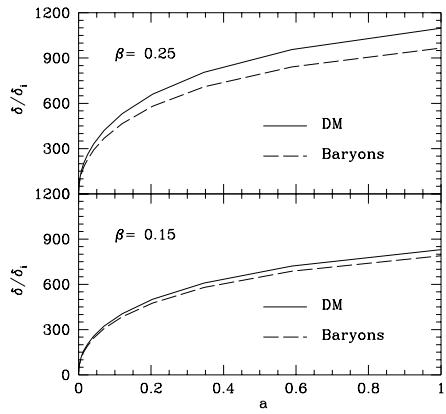}
\caption{DM and baryons linear perturbation growth for two different values
of $\beta $. The dependence on $\alpha $ is weak and could not be
appreciated in this plot. \protect \protect \\}

\label{fig:b1}
\end{figure}

From eqs.~(\ref{delta2},\ref{delta2b}), the linear evolution of the density perturbations
can be easily worked out (in some cases \cite{amen99} this can be
done analytically). In Fig.~\ref{fig:b1} we show $\delta ^{c,b}$
as a function of the scale factor $a$. 

\begin{figure}
\includegraphics[  scale=0.3]{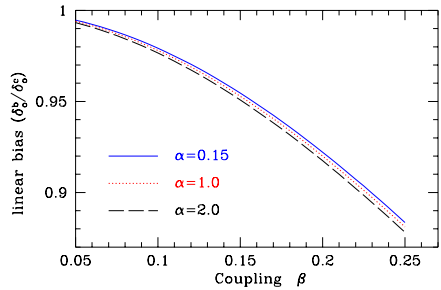}
\caption{Linear bias as a function of $\beta $ for three values of $\alpha $.
Notice the very weak dependence on $\alpha $.}

\label{fig:b2}
\end{figure}

\begin{figure}
\includegraphics[  scale=0.3]{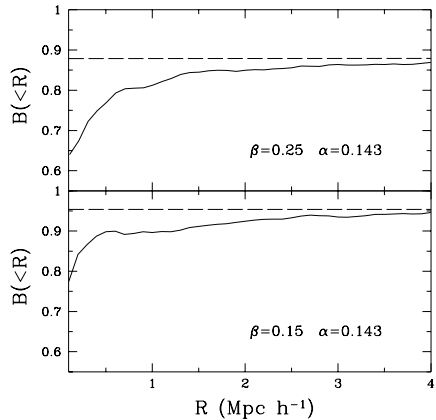}

\caption{Behaviour of the integrated bias $B$ for $\beta =0.15$ and for
$\beta =0.25$. Notice that $B$ tends to the predicted linear bias
(dashed horizontal lines) at large scales.}
\label{fig:b3}
\end{figure}

As a consequence of these dynamical equations, $\delta ^{c}$ develops
a bias with respect to $\delta ^{b}$, due to the extra gravity felt
by DM. At the present epoch, this bias, found in the linear theory,
is well fitted by the following empirical expression: \begin{equation}
b(\alpha ,\beta )=\frac{\delta _{o}^{b}}{\delta _{o}^{c}}=\frac{1}{1+0.015\, \alpha \beta +2.1\, \beta ^{2}}.\label{lin-bias}\end{equation}
 Both this expression and Fig.~\ref{fig:b2} show that the bias $b$
depends on $\beta $, while its dependence on $\alpha $ is very weak.
Using the high resolution clusters we can test the behavior of the
bias in the highly non--linear regime. To do so, we define the integrated
bias $B$ as: \[
B(<r)=\frac{\rho _{b}(<r)-\hat{\rho }_{b}}{\hat{\rho }_{b}}\cdot \frac{\hat{\rho _{c}}}{\rho _{c}(<r)-\hat{\rho }_{c}},\]
 where $\rho _{c}(<r)$ and $\rho _{b}(<r)$ are calculated inside
a radius $r$ from the halo center and we use a hat (~$\hat{}$~)
to denote average densities. In order to avoid problems with force
resolution, the central zone ($r<10h^{-1}$kpc) of the halo is not
used. In Fig.~\ref{fig:b3} we show $B(<r)$ for the same halo, in
cosmologies characterized by different coupling parameters $\beta $,
keeping all other parameters equal. Fig.~\ref{fig:b3} shows that
non--linearity significantly enhances the expected bias; however,
at larges scales, we recover the theoretical linear value, as provided
by eq.~(\ref{lin-bias}). 

\begin{figure}
\includegraphics[  scale=0.3]{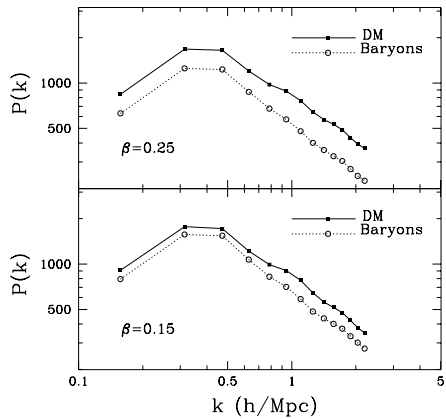}
\caption{Power spectra for DM and baryon particles evaluated from the simulations
averaging over 60 random observers. The increase of the bias at small
scales appears clearly.}

\label{fig:pk}
\end{figure}

The scale dependence of bias can also be appreciated from Fig.~\ref{fig:pk},
where power spectra for baryons and DM, worked out from simulations
at $z=0$, are shown. 

\subsection{Density profiles}

\label{sec:density} Let us remind again that all simulations are
started from the same random numbers. Therefore, it comes as no surprise
that they yield similar world pictures. In the $\Lambda $CDM simulation,
we selected a halo, whose \textit{virial} radius $r_{v}=812\, h^{-1}$kpc
encloses a mass $M_{v}=6.45\cdot 10^{13}h^{-1}M_{\odot }$. Similar
halos, located in the same place, are set in all other models considered.
We then run new simulations of all models in Table 1, magnifying the
region centered on this halo. To do so, short waves were first added
to the initial perturbation spectrum in all simulations. 

In $\Lambda $CDM, the halo profile is accurately fitted by a NFW
expression \cite{nfw96,nfw97}: \[
{\frac{\rho (r)}{\rho _{cr}}}={\frac{\delta ^{*}}{{\frac{r}{r_{s}}}\big (1+{\frac{r}{r_{s}}}\big )^{2}}}\]
 with a scale radius $r_{s}=0.249\, h^{-1}$Mpc (here $\rho _{cr}$
is the critical density and $\delta ^{*}$ is a parameter which sets
the halo density contrast). 

When the same halo is magnified in coupled DE models, we find model
dependent behaviors towards the halo center. The essential restrictions
to coupled DE models, arising from the non--linear treatment, derive
from these behaviors. However, in spite of such model dependence in
the central areas, the outer parts of halos ($R>100\, h^{-1}$Mpc)
show discrepancies that, from $100\, h^{-1}$kpc to $700h^{-1}$kpc,
never exceed $\sim 10\, \%$. 

\begin{figure}
\includegraphics[  scale=0.3]{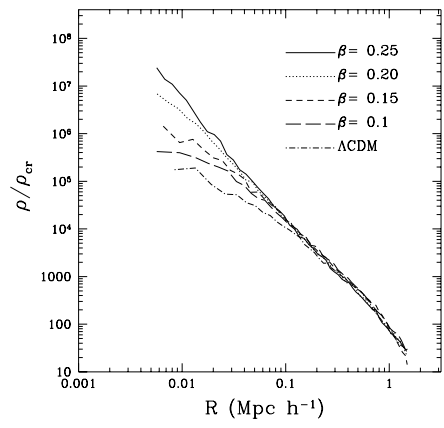}
\caption{DM density profile for four different coupling values and for $\Lambda $CDM.}

\label{fig:prof}
\end{figure}

\begin{figure}
\includegraphics[  scale=0.3]{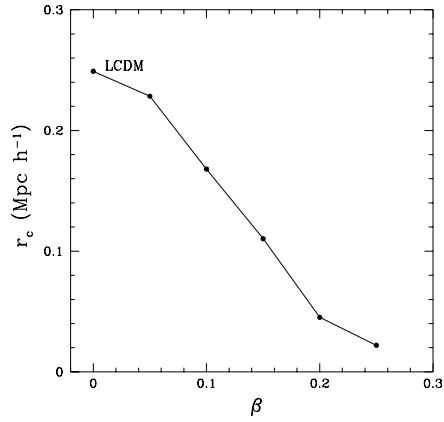}
\caption{Scale radius of a NFW profile as a function of the coupling parameter
$\beta $.}

\label{fig:rs}
\end{figure}

Let us now discuss the substantial model dependence found in the central
region ($R<100\, h^{-1}$kpc). It was already known that halos are
denser in dynamical DE than in $\Lambda $CDM \cite{kmmb}, although
the density enhancement is fairly small and hardly exceeds $\sim 40\, \%$.
Higher density means smaller $r_{s}$. The coupled DE simulations
we perfomed show that the dark--dark coupling tends to enhance such
effect. In Fig.~\ref{fig:prof} we overlap the profiles of the DM
components of all our models, starting from $\Lambda CDM$ (lower
curve), up to a RP model with coupling parameter $\beta =0.25$ (upper
curve). The values of $r_{s}$ change from $\simeq 0.25\, h^{-1}$Mpc
($\Lambda CDM$) to $\simeq 0.022\, h^{-1}$Mpc ($\beta =0.25$).
The dependence of $r_{s}$ on $\beta $ is plotted in Fig.~\ref{fig:rs}. 

\begin{figure}
\includegraphics[  scale=0.3]{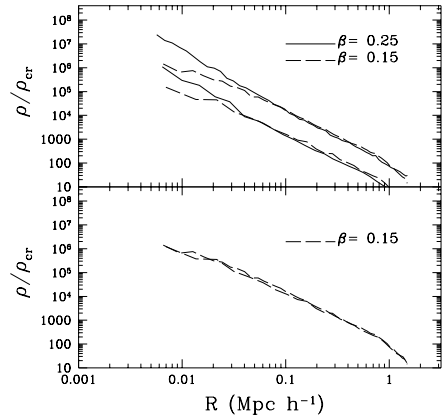}
\caption{Upper panel: DM and baryons density profiles (respectively upper
curve and lower curve) for $\beta =0.15$ and for $\beta =0.25$.
Lower panel: once rescaled taking into account the different values
$\Omega _{b}$ and $\Omega _{c}$, there is no discrepancy between
DM and baryons.}

\label{fig:prof2}
\end{figure}

\begin{figure}
\includegraphics[  scale=0.3]{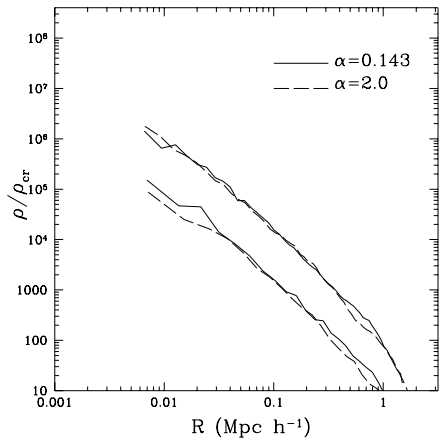}
\caption{DM and baryons density profiles for $\alpha =2.$ and $\alpha =0.143$
(here $\beta =0.15$).}

\label{fig:prof3}
\end{figure}

In order to make sure that this effect was not related to some peculiarity
of the halo selected, we magnified two other halos of a simulation
with $\beta =0.25$. Here we found even lower values for the scale
radius $r_{s}$ ($0.0105\, h^{-1}$Mpc and $0.0103\, h^{-1}$Mpc respectively). 

As a matter of fact, the profiles found $\beta =0.25$ or $0.2$ can
be fitted by a single power law: \[
\frac{\rho (r)}{\rho _{cr}}\propto r^{\gamma }\]
 in the whole dynamical range, i.e., from $r=1.0h^{-1}$Mpc down to
$r=0.005h^{-1}$Mpc (resolution limit), with a value of $\gamma \simeq -2.30$. 

An analysis of Fig.~\ref{fig:rs} shows that, only for $\beta $
as low as $\simeq 0.1$, $r_{s}$ attains half the value for $\Lambda $CDM.
Accordingly, we may consider viable coupled RP models only with $\beta <0.1$. 

Simulations distinguish baryons from DM particles, as already discussed
in the bias subsection. This allows us to draw separate density profiles.
They are shown in Fig.~\ref{fig:prof2}, for two different coupled
DE models (upper panel). No apparent discrepancy between DM and baryon
profiles can be seen: they overlap fairly well, once we rescale them
taking into account the different values of $\Omega {_{b}}$ and $\Omega {_{c}}$
(Fig~\ref{fig:prof2} lower panel). 

In Fig.~\ref{fig:prof3} we compare the profiles of the same halo,
with two different values of $\alpha $ (2.0 and 0.143), but with
the same coupling ($\beta =0.15$). The profiles overlap very well
both for DM particles (upper curves) and for baryons (lower curves).
We conclude that the slope of profiles depends very weekly on $\alpha $.

\section{Conclusions}

\label{sec:conclusions} After finding that coupled DE models are
consistent with those observables whose behavior can be predicted
at the linear level \cite{amen02,am-mnras,amed02}, a test of their
non--linear behavior had to be carried on. An optimistic hope was
that coupled DE models helped to solve some of the long standing contradictions
between observations and theoretical or numerical predictions (e.g. \cite{moore}). In
particular, one could hope to find halo profiles whose shape is not
NFW (if this is still a problem) with a slope distribution closer
to the observed ones for low surface brightness galaxies \cite{deb,swa} and
spiral galaxies \cite{salucci}. From this point of view, coupled
DE with a RP potential leads to modest results. Very high coupling
levels, instead of producing a flatter core, yield profiles still
farther from observations. In all cases, the problem with concentration
distribution is not solved, just as when making recourse to models
with (uncoupled) dynamical DE. 

It should be also reminded that the RP potential considered here is
characterized by very low $\alpha $ values. The scale factor dependence
of the state parameter $w$, for such values of $\alpha $, is shown
in Fig.~\ref{fig:den} and approaches -1 already at redshifts $\sim 3$--4.
As far as the state parameter is concerned, therefore, these models
are quite close to $\Lambda $CDM and, in a sense, could be considered
a variant of $\Lambda $CDM models, for which DE is coupled to DM. 

In spite of the lack of improvement for what concerns slopes, N--body
simulations lead to really significant results. First of all, the
parameter space for coupled DE models is restricted to couplings $\beta <0.1$,
however leaving a wide room for significant couplings. Apart of the
question of profiles, the halo mass function has been tested and found
consistent with other DE models and with observations. Its evolution
has been predicted and can be tested against future data. From this
point of view, therefore, coupled DE passed the non--linear test. 

Performing N--body simulations was also important to test the evolution
of the \textit{bias}, between baryon and DM fluctuations, which is
one of the main characteristics of coupled DE models at the linear
level. Here we showed that the bias still exists, and actually increases,
in the non--linear structures and studied its evolution. These results
are open to an accurate comparison with data, that shall be deepened
elsewhere. 

Let us however conclude these comments with a further observation.
Coupled DE apparently leads to higher halo concentration essentially
because of the evolution of the mass of DM particles and of the coupling
constant between them. In the simulations we run, such mass depends
on time and gradually decreases, as is predicted by coupled DE theories
at the Newtonian approximation level. Accordingly, each DM particle
mass was greater than today, in the past. Its gravity was therefore
stronger. This is the reason why, although normalizing all models
to the same $\sigma _{8}$ at $z=0$, we produce more concentrated
halos: the forces which bound them were stronger in the past than
today. 

After appreciating this point, one can tentatively propose a way out
for the halo concentration problem: a coupled DE model leading to
DM particle masses which increase in time. This increase should also
be fast enough to beat the higher gravity constant binding DM particels.
This takes us back to the selection of a suitable self--interaction
potential $V(\phi )$, which has no immediate obvious solution. This
problem shall be therefore deepened in future work.

\end{document}